Paper ID #

# Cloud-based traffic data fusion for situation evaluation of handover scenarios

Andreas Otte[1*], Jens Staub[1], Jonas Vogt[1], Horst Wieker[1]

1. htw saar, University of Applied Sciences Saarbrücken, Germany, andreas.otte@htwsaar.de

**Abstract**

Upcoming vehicles introduce functions at the level of conditional automation where a driver no longer must supervise the system but must be able to take over the driving function when the system request it. This leads to the situation that the driver does not concentrate on the road but is reading mails for example. In this case, the driver is not able to take over the driving function immediately because she must first orient herself in the current traffic situation. In an urban scenario a situation that an automated vehicle is not able to steer further can arise quickly. To find suitable handover situations, data from traffic infrastructure systems, vehicles, and drivers is fused in a cloud-based situation to provide the hole traffic environment as base for the decision when the driving function should be transferred best and possibly even before a critical situation arises.

**Keywords:**
TRAFFIC DATA FUSION and AUTOMATED DRIVING HANDOVER

**Introduction**

Modern road infrastructure is already able to capture information about traffic situations. Today, sensors like traffic cameras, induction loops, or floating car data [1] are mostly used to determine traffic flow and road utilization. The use of optical sensors like camera systems not only allows the detection of road users, but also the determination of their location, classification, direction of movement and speed. This information is especially valuable for automated vehicles: they can store this information in their Local Dynamic Map (LDM) [2] and use it for their trajectories planning and for collision avoidance calculations. Furthermore, in conditionally automated vehicles of SAE level three [3], this information can also help to plan and execute handover manoeuvres between the automated driving function and the human driver. These manoeuvres require the driver to take over control of the vehicle and are necessary if the automated vehicle is not capable to handle a current situation or a situation closely ahead. By using additional sources of information, this handover process can be performed more safely and foresighted.

The German research project kantSaar [4,5] aims to capture the driver's state by using sensors inside of the vehicle. The in-vehicle sensors track information about the vehicle's state and trajectory, as well as the driver's level of attention and stress. Fusing the neurocognitive data of the driver, the vehicle's state and the traffic and environmental information from the vehicle and the traffic infrastructure to a combined data set creates a model of the current traffic situation. The set combines data subjectively experienced by the driver and data objectively perceived by the sensors. Up until now, infrastructure data fusion for intelligent transportation systems is mostly done only from one side, infrastructure or vehicle (e.g. [6,7,8,9]). Jabber et al. [10] are working on a similar approach as kantSaar but only based on data from the vehicle and no data from the traffic infrastructure or V2X data. Additionally, they use an existing data set for the in-vehicle driver information whereas kantSaar creates own passenger data to be able to create a minimal intrusive system. A database with such traffic situations can be used to identify and evaluate stressful situations and draw conclusions about the effect of specific parameters on the vehicle and its driver in specific situation. This knowledge allows the creation of a map, showing road segments, either difficult for automated vehicles or stressful for the driver. In the future, an AI-based generalization and transfer of the results to other cities and intersections is planned, which should minimize the sensor data needed and make monitoring of every driver redundant.

**Approach**

The neurocognitive data and the traffic data originate from two different (sub-) systems. In a first step, the data needs to be gathered. As traffic is a highly dynamic environment, it cannot be ensured that a single local node can collect all information about the current traffic situation. Cooperative vehicles exchange information about their position, speed and heading by using Cooperative Awareness Messages (CAM) [11]. In turn, road infrastructure can detect non-cooperative vehicles and vulnerable road users (VRU) by using sensors and detectors. The information about their position, speed, direction of movement etc. can be transmitted using Collective Perception Messages (CPM) [12]. Those CAM and CPM are transmitted using dedicated short-range communications (DSRC) like ETSI ITS-G5 or C-V2X. The vehicle under test (VUT) as part of the DSRC network cannot ensure to receive every message and has therefore a possible lack of information. This also applies to the road infrastructure, which is in addition not aware of a driver's condition. Therefore, none of the participants possess a complete set of information and is only able to perform an incomplete data fusion. To fuse all the gathered data, a cloud service is introduced.

Before the data can be transmitted to the cloud-based Data Fusion service (DF), it needs to be supplemented with meta information. Time and location information as well as the identity of the data originator need to be added to the data sets. Positioning information originates from global navigation satellite systems like GPS. GPS also provides timing information and allows to synchronize the participants of the DSRC network. The originating system is identified by a temporary identifier, e.g., the station ID of the ITS facility layer. The data originating stations need a connection to the cloud-based data fusion service. In case of the VUT, cellular communication is used to transfer the collected data to the DF. Regarding differing coverage, the transmission shall be able to be performed with low signal quality. Therefore, the sensor data and its meta information are packed sparingly. To reduce the overhead of reliable transmission, the data is collected over a time interval and is then transmitted with a single transaction. Within each transmission, a meta information block with an absolute reference time, position and identity is created. Every individual information contains a relative time and position, which is smaller than the absolute values. To enable interoperability, the data needs to be serialized. The

Cloud-based traffic data fusion for situation evaluation of handover scenarios

DF is then independent of the implementation of the data aggregators. It receives the collected data and stores it in a database. First, each type of information, for instance CAM information, CPM information or the sensor set of the VUT and the infrastructure, gets stored in separate tables. Then, the Data Fusion checks for and if found deletes duplicates in each information table. The DF processes the data, select road users from the different sets of raw data, merges them and creates a digital map. Therefore, the data previously contained in multiple information tables, gets joined into one data structure and duplicate entries, for example emerging from the simultaneous detection of an object via multiple different ways, get eliminated. The process described here has not been designed to perform an in situ analysis and to directly send a recommendation to a partially automated vehicle. The goal is rather to evaluate situations later, to be able to draw conclusions for similar situations in the future without the need of cloud-based data processing. This approach is therefore independent of transmission duration by the used cellular connection and the data can be periodically transmitted in batches.

*Architecture*

The system architecture of the data aggregators and the Data Fusion is shown in Figure 1. It is split in two parts: The Remote Station Plane and the Backend Plane. The Remote Station Plane contains infrastructure systems called roadside ITS stations (ITS-S) and vehicle ITS-S. The VUT is a research vehicle ITS-S with a driver monitoring system. The backend plane consists of services pre-processing gathered data of the remote station plane and additional third-party data suppliers as well as the Data Fusion and the Situation Evaluation. The raw, fused and evaluated data is stored in the Situation Storage.

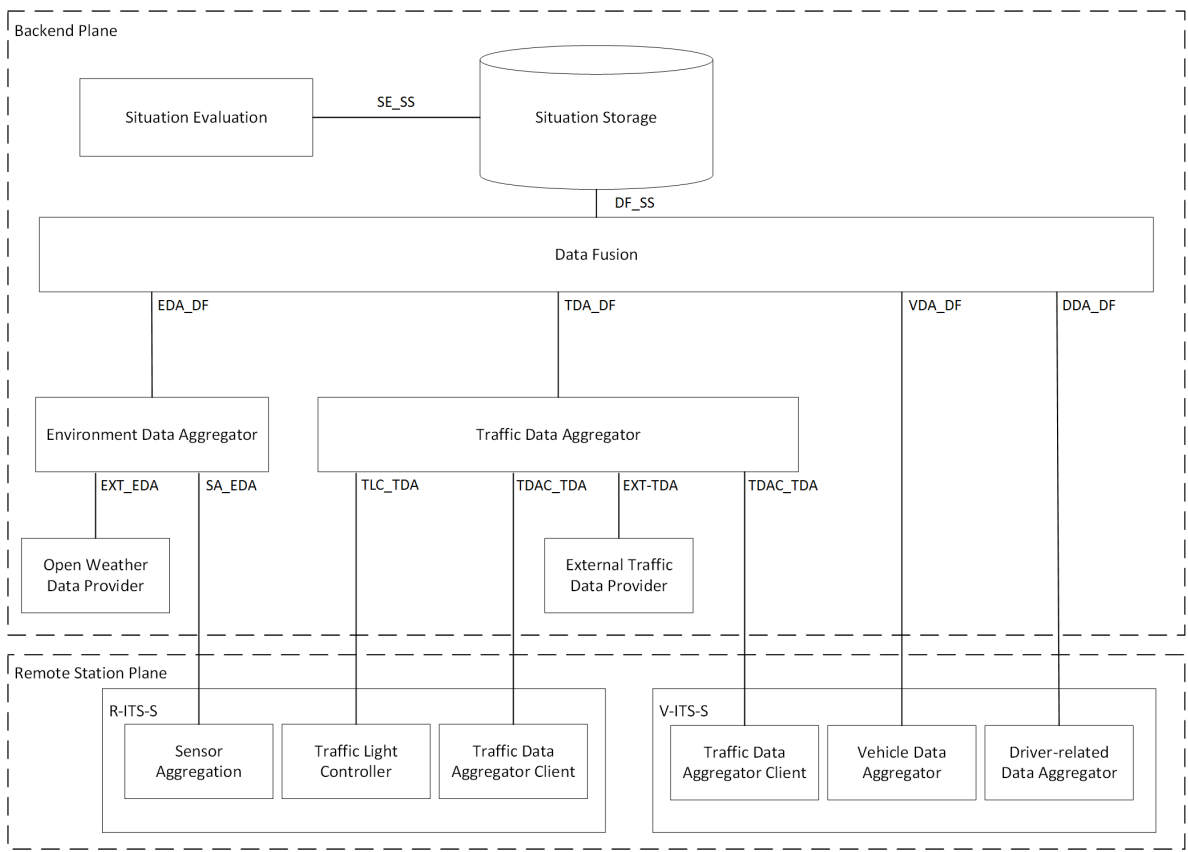

**Figure 1 Aggregation and fusion architecture**

Cloud-based traffic data fusion for situation evaluation of handover scenarios

The Traffic Data Aggregator Client (TDAC) is present in both vehicle and infrastructure system. Its purpose is the collection of V2X messages that originated from or were received by this station. Especially Cooperative Awareness Messages (CAM), Decentralized Environmental Notification Messages (DENM), Signal Phase and Timing Messages (SPAT) and Collective Perception Messages (CPM) are gathered. MAP Messages describing the topology of an intersection are not transmitted but already stored in the Traffic Data Aggregator, as they change rarely. From these message types, only a subset of valuable information is gathered, respectively. Further information is gathered from road infrastructure, which can be equipped with various sensors. Modern traffic light systems react on changing traffic conditions and switch signal phases dynamically. The information about different triggers like induction loops, optical sensors or specialized prioritization systems for public transport can be gathered as well and sent to the Traffic Data Aggregator on the backend plane.

Environmental sensors, for example to measure air quality, air and surface temperature, or pollution and noise level can be attached to roadside ITS-S and supplement the sensor data. Those data are sent to the Environment Data Aggregator in the backend plane.

The Vehicle Data Aggregator (VDA) collects sensor data of the VUT directly from the vehicle's CAN bus systems. To allow an efficient use of bandwidth, the different types of CAN data are transmitted in different temporal intervals, for instance the current lateral acceleration is transmitted more often as the state of the wiper system. Using a sensor data abstraction layer, so the same implementation of the aggregator can be used in different VUT, the VDA registers for a subset of the available vehicle data listed in Table 1.

Table 1 Selection of in-vehicle sensor values

| Value | Description |
|---|---|
| **Brake System** | Status of the brake system including brake actuation, status of Antilock Braking System and if a panic braking is performed |
| **Clutch and Gear status** | Status of the clutch pedal (if available) and the current gear |
| **Door position** | Position of each door (closed, ajar or open) |
| **Exterior Lights** | Low and High beam status, fog lights, emergency lights, hazard warning system, turning signals |
| **GNSS** | GNNS position information |
| **Speed** | Vehicle speed and longitudinal and latitudinal acceleration |
| **Rain Sensor** | Rain intensity sensor and wiper system |
| **Yaw** | Yaw rate, yaw velocity, steering wheel angle and steering wheel velocity |

As the transmission of datasets involves overhead, the sensor data is not directly forwarded on reception but collected and stored in a local vehicle data storage. Therefore, the collected data is transmitted periodically via the interface VDA_DF. The period can be adapted on changing circumstances (mobile reception quality, number of changes, etc.). If the transmission was successfully acknowledged, the local storage is cleared.

The Driver-related Data Aggregator (DDA) has access to sensors that monitor the state of a passenger. A combination of contact-based and contactless sensors tracks heart rate, surface temperature in specific regions of the driver's face, skin conduction and brain activity (EEG). Based on this data, the level of stress, fatigue or even distraction can be determined [13]. Stress is further stripped down to five-scaled arousal and valence values following the Self-Assessment Manikin (SAM) [14] method. With the use of SAM, a later self-evaluation of the driver's emotional state as well as the automated capturing by the mentioned sensors is enabled.

The function of the Environment Data Aggregator is to collect environment data and provide it to the Data Fusion. This data is primarily weather data. Weather information is an important part of road traffic and can significantly affect a traffic situation. The weather data, which is collected in the Environment Data Aggregator, is bound to a timestamp with a validity as well as a location area. It describes the current weather condition using information about temperature, precipitation, wind, light and visibility conditions, pressure, humidity, and cloudiness.

The Traffic Data Aggregator receives information from various TDAC. Its purpose is to preprocess the information. As V2X messages can be received by different participating stations in an area, its containing information can be forwarded more than once. The Traffic Data Aggregator identifies and deletes those duplicates. The resulting set of messages is then sent to the Data Fusion.

The Data Fusion is the central component of the backend plane. It receives all defined data classes: Environmental data, traffic data, vehicle data and driver-related data. On reception, the Data Fusion stores the raw data in the Situation Database. The goal of the Data Fusion is a digital representation of traffic situations merged with information from the environment, the VUT and the driver.

*Data Fusion Process*

To create traffic situations, the data fusion process is triggered manually on request by an operator by choosing the VUT identifier and the desired timestamp. In a first step, the data sets to fuse are queried with the use of time and location information given by the requested timestamp and the location of the VUT. In order to eliminate duplicated information within the traffic objects provided by various sources, the traffic objects need to be transformed to a common data format. Duplicates may occur while joining CAM and CPM information as well as the position vector of the VUT and the infrastructure sensor set. The quality of the fusion depends on the information quality of the detected traffic objects. A flawless accuracy cannot be reached in real environments, which leads to unequal digital representations of the same real traffic object. Only the similarity between the objects can be used to identify objects representing the same real object. Two objects are similar, if the differences regarding their positions, directions of movement, speeds and classifications are within defined thresholds that are chosen by the estimated error of the input sensors. The traffic objects are clustered using dynamic course ranges depending on the objects speed. A similarity check is performed within the clusters to identify and remove duplicates. Clustering by courses provided a performance advantage over clustering based on the location of the objects when focusing on intersections as the number of comparisons could be reduced. Focusing on highways or rural roads, this clustering method does not provide any advantage.

Another aim of the data fusion process is to provide a structured access for later evaluation. Therefore, a relational database is used. The data base scheme is provided in Figure 2. A situation is described by a geographical area and a point in time as well as a unique situation identifier. All matching detected objects extracted from CAM, CPM and other sensors are merged and again checked for duplicates. Topology information gathered from MAP messages is directly merged with the signal phase information stored in corresponding SPAT messages. The sensor data of the VUT and the driver-related data do not differ to its raw format and therefore just needs to be copied and linked to the situation. Also, hazardous events, like a panic braking or an emergency vehicle warning are stored and linked to the situation.

Every data, matching position and time, is linked to the ID of the situation. Additionally, traffic objects can be linked to a lane, to create a lane-wise representation of a road segment. Eventually the evaluation of the recorded situations is performed and the suitability for a handover is determined.

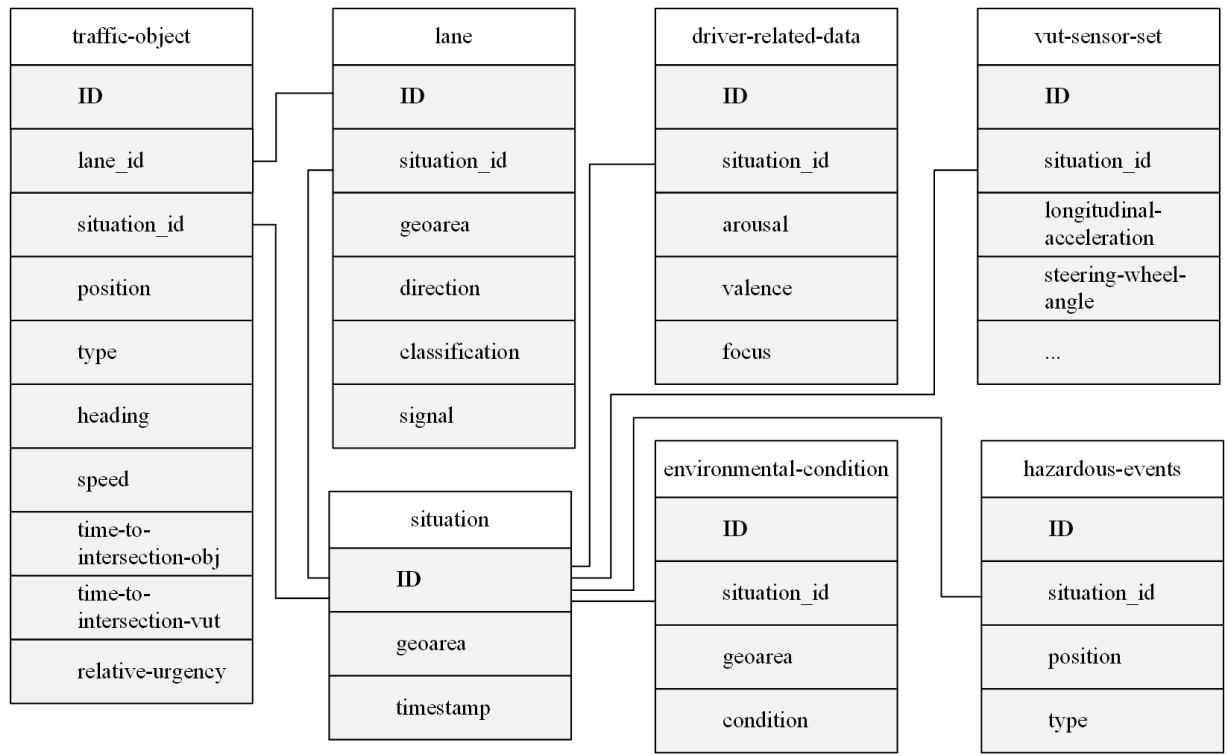

Figure 2 Simplified database scheme of situation storage

**Results**

The described aggregation and fusion architecture has been used in test drives in the German cities of Saarbruecken and Merzig. Within twelve months, about 6.000 situations have been created by the data fusion based on 2.0 million camera-based detected objects (as parts of the CPMs), 3.8 million Cooperative-Awareness Messages (CAM) and 600.000 sensor extracts of the vehicle under test.

Cloud-based traffic data fusion for situation evaluation of handover scenarios

Additionally, 2.500 records of the driver status were captured.

Figure 3 shows an actual created situation visualized by a map application. The blue car represents the VUT, the green cars and the pedestrians were detected by the infrastructure camera system and sensors of the VUT including V2X technology. For the map representations in the example shown, map tiles from OpenStreetMap [15] and CARTO [16] were used.

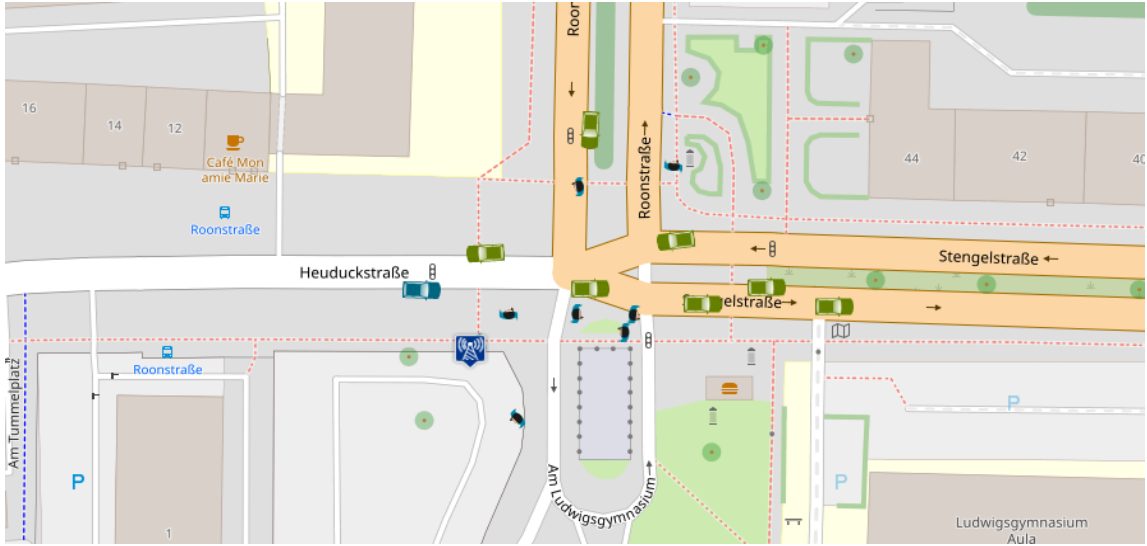

**Figure 3 Visualization of a recorded situation**

In addition to the visual representation on the map, Table 2 contains the position vector of each detected dynamic traffic object as well as four parameters for later evaluation. Besides the *distance* of the traffic object to the VUT, two different *time to intersection (TTI)* and a *relative urgency (RU)* is provided. $TTI_{OBJ}$ describes how many milliseconds a traffic object takes to get to the intersection point with the VUT, $TTI_{VUT}$ accordingly. If TTI is set to *-1*, there is no intersection between the traffic object and the VUT in the future. RU provides an estimate of how fast two objects shorten their distance to each other. The higher the RU value, the smaller the approach.

**Table 2 Traffic objects of the resulting situation**

| ID | Classification | Lat (°) | Lon (°) | Speed (m/s) | Course (°) | Distance (m) | $TTI_{OBJ}$ (ms) | $TTI_{VUT}$ (ms) | RU |
|---|---|---|---|---|---|---|---|---|---|
| 3 | PASSENGER CAR | 49.2340183 | 6.9823835 | 3.46 | 267.2 | 11.27 | -1 | -1 | 893 |
| 2 | PASSENGER CAR | 49.2339679 | 6.9826028 | 7.57 | 89.0 | 25.62 | -1 | -1 | 6598 |
| 47 | PEDESTRIAN | 49.2339251 | 6.9826933 | 1.48 | 270.8 | 32.58 | -1 | -1 | 2530 |
| 10 | PASSENGER CAR | 49.2339447 | 6.983114 | 14.22 | 90.0 | 62.78 | -1 | -1 | MAX |
| 188 | PASSENGER CAR | 49.2340355 | 6.9827828 | 3.18 | 263.6 | 39.44 | -1 | -1 | 2759 |
| 143 | PEDESTRIAN | 49.2338076 | 6.9824494 | 0.41 | 225.0 | 22.85 | -1 | -1 | 2980 |
| 58 | PASSENGER CAR | 49.2339706 | 6.982973 | 8.13 | 89.1 | 52.50 | -1 | -1 | 15811 |
| 201 | PASSENGER CAR | 49.2339473 | 6.9828387 | 10.33 | 90.0 | 42.80 | -1 | -1 | 37118 |
| 160 | PEDESTRIAN | 49.2341248 | 6.9827959 | 4.04 | 3.9 | 43.36 | -1 | -1 | 5194 |

Cloud-based traffic data fusion for situation evaluation of handover scenarios

| 149 | PEDESTRIAN | 49.2339105 | 6.9823796 | 1.69 | 243.4 | 11.29 | -1 | -1 | 1054 |
| 200 | PASSENGER CAR | 49.2341879 | 6.9826028 | 7.57 | 183.0 | 35.52 | 3720 | 3089 | 2654 |
| 481 | PEDESTRIAN | 49.2339219 | 6.9826735 | 1.48 | 95.6 | 31.15 | -1 | -1 | 3138 |
| 48 | PEDESTRIAN | 49.2339224 | 6.9824532 | 2.57 | 356.0 | 15.59 | 1639 | 1315 | 1295 |
| 162 | PEDESTRIAN | 49.2341248 | 6.9825959 | 1.86 | 260.3 | 30.66 | -1 | -1 | 2857 |

Driver-related data in form of stress stripped in valence and arousal are gathered as well. Valence and arousal can each take values on a five-point scale. A *one* as an arousal value means exited or frenzied whereas a *five* means calm or sleepy. A *one* as a valence value means happy or pleased whereas a *five* means unhappy or melancholic. A *three* in both scales corresponds to a neutral stress level. In this very example, the driver performed a self-evaluation and rated a *two* in valence and a *three* in arousal, which a not very stressful situation. Based on the gathered stress data of the driver along the test track, a stress-map was created as shown in Figure 4.

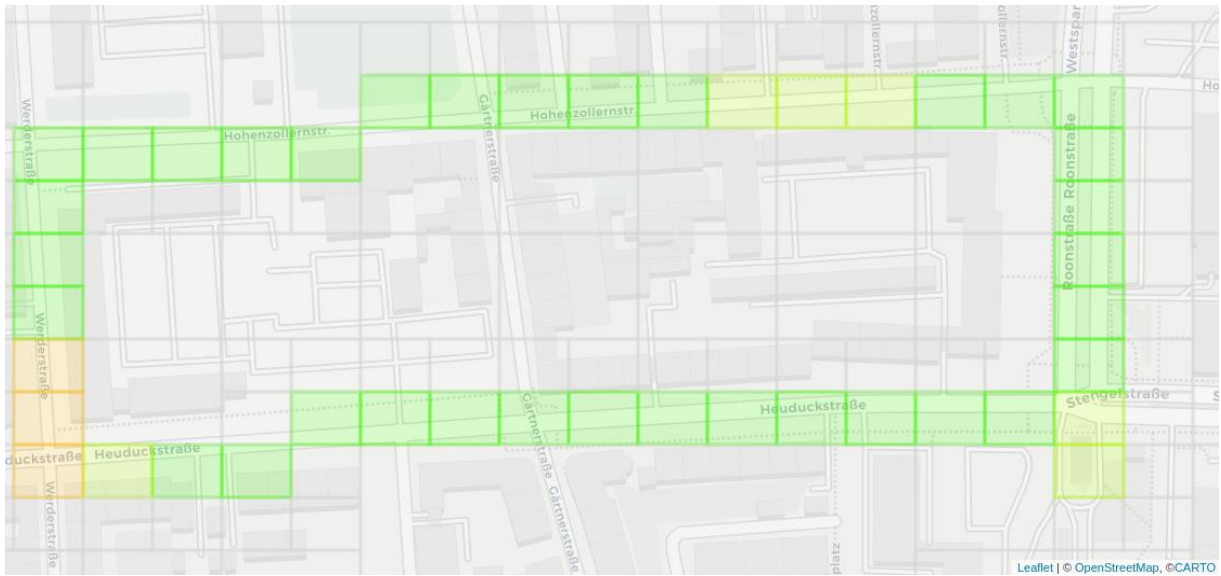

**Figure 4 Stress-map of the test drive in the city of Saarbruecken**

The stress-map utilize a quad tree [17] to store and combine the stress information along the test track. The used colour code is based on the exemplary matrix shown in Figure 5. The exact interpretation of the valence and arousal values is to be done in a later step and is not part of this paper.

Cloud-based traffic data fusion for situation evaluation of handover scenarios

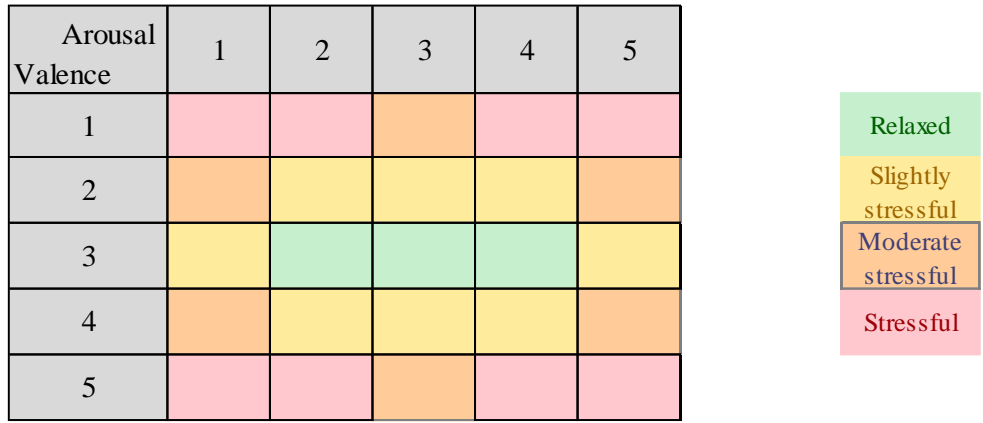

Figure 5 Exemplary matrix for the interpretation of the level of stress

**Conclusion**

The project results have shown that the previously described architecture has been implemented and used successfully. The database of traffic situations, generated during the numerous test drives, in addition to the driver-related data, creates a new dimension for evaluating traffic situations, by accounting the drivers levels of stress and attention. This evaluation can be performed in a traffic control centre to gain and maybe even provide information about current traffic. Furthermore, the data sets can be used to train neural networks, to forecast stressful traffic situations and provide information for useful handover manoeuvres between the driver and the automated driving function.

Additionally, our research has shown, that monitoring multiple VUT in the same traffic situation might give further insight on the complexity and stressfulness of this situation, by providing multiple driver perspectives. This is a topic for future research.

**Acknowledgements**

The work of this paper has been funded by the German Federal Ministry of Transport and Digital Infrastructure within the project kantSaar (grant number 16AVF2129). The operative project consortium consists of the University of Applied Sciences Saarland and the University of Saarland. The project approach and the partial project outcomes result from the collaborative work of the entire kantSaar project team of htw saar.